\documentclass[journal]{IEEEtran}

\pdfoutput=1
\usepackage[utf8]{inputenc}
\usepackage[english]{babel}
\usepackage[T1]{fontenc}
\usepackage{float}
\usepackage{tikz}
\usepackage{graphicx}
\usepackage{mathptmx, helvet, courier}
\usepackage{amsmath,amssymb}
\usepackage{array}
\usepackage[fixlanguage]{babelbib}
\selectbiblanguage{english}
\usepackage[colorlinks,linkcolor=black,bookmarks=false]{hyperref}
\graphicspath{{img/}}
\DeclareGraphicsExtensions{.pdf,.png,.jpg,.JPG}
\newcolumntype{C}[1]{>{\centering\arraybackslash}m{#1}}

\newcommand{\thauthor}{\IEEEauthorblockN{Dmytro Levit, Igor Konorov, Daniel Greenwald, Stephan Paul} \\
\IEEEauthorblockA{Technische Universität München
}
}

\newcommand{\thtitle}{FPGA Based Data Read-Out System of the Belle 2 Pixel Detector}

\begin{document}
\bstctlcite{IEEEexample:BSTcontrol}

\author{\thauthor}
\title{\thtitle}

\maketitle
\thispagestyle{empty}
\pagestyle{empty}

\begin{abstract}
	The upgrades of the Belle experiment and the KEKB accelerator aim to increase the data set of the experiment by the factor 50.
	This will be achieved by increasing the luminosity of the accelerator which requires a significant upgrade of the detector.
	A new pixel detector based on DEPFET technology will be installed to handle the increased reaction rate and provide better vertex resolution.
	One of the features of the DEPFET detector is a long integration time of 20\,$\mu$s, which increases detector occupancy up to 3\,\%.
	The detector will generate about 2\,GB/s of data.
	An FPGA-based two-level read-out system, the Data Handling Hybrid, was developed for the Belle 2 pixel detector.
	The system consists of 40 read-out and 8 controller modules.
	All modules are built in $\mu$TCA form factor using Xilinx \mbox{Virtex-6} FPGA and can utilize up to 4\,GB DDR3 RAM.
	The system was successfully tested in the beam test at DESY in January 2014.
	The functionality and the architecture of the Belle 2 Data Handling Hybrid system as well as the performance of the system during the beam test are presented in the paper.

\end{abstract}

\section{Introduction}

The upgrade of the Belle experiment aims to increase the recorded data set of the B Factory by factor 50.
Besides the upgrade of the accelerator SuperKEKB\,\cite{Ohnishi01032013} that will bring an increase in the luminosity by adopting a nano-beam collision scheme at the interaction point, the significant upgrade of all detectors of the experiment is required to cope with the increased reaction rate.
The details of the upgrades can be found elsewhere\,\cite{b2tdr}. 

\section{Pixel Detector}

The silicon pixel detector will be introduced as part of the Belle 2 silicon vertex detector, to be the innermost detector layer.
The silicon pixel detector is an active-pixel detector built using DEPFET\,\cite{Kemmer1987365} technology.
This technology allows us to build a detector with a very low material budget\,(0.19\,\%\,$X_0$ per layer\,\cite{1748-0221-7-01-C01014}), which reduces multiple scattering and provides spatial resolution below 10\,$\mu$m, improving vertexing.

The matrix of a pixel detector module, called a half ladder, consists of 768x250 DEPFET pixels.
The integration time of a half ladder is 20\,$\mu$s.
Two half ladders are glued on the far edge of the silicon frame together and form a mechanical module, a ladder.
The 20 mechanical modules of the pixel detector are arranged in two cylindrical layers around the interaction point of the accelerator.
The inner layer consists of 8 modules with an average radius of 14\,mm and sensitive length of 90\,mm.
The outer layer consists of 12 modules with a radius of 22\,mm and sensitive length of 123\,mm\,\cite{1748-0221-7-01-C01014}.

	\begin{figure}
		\includegraphics[width=3.5in]{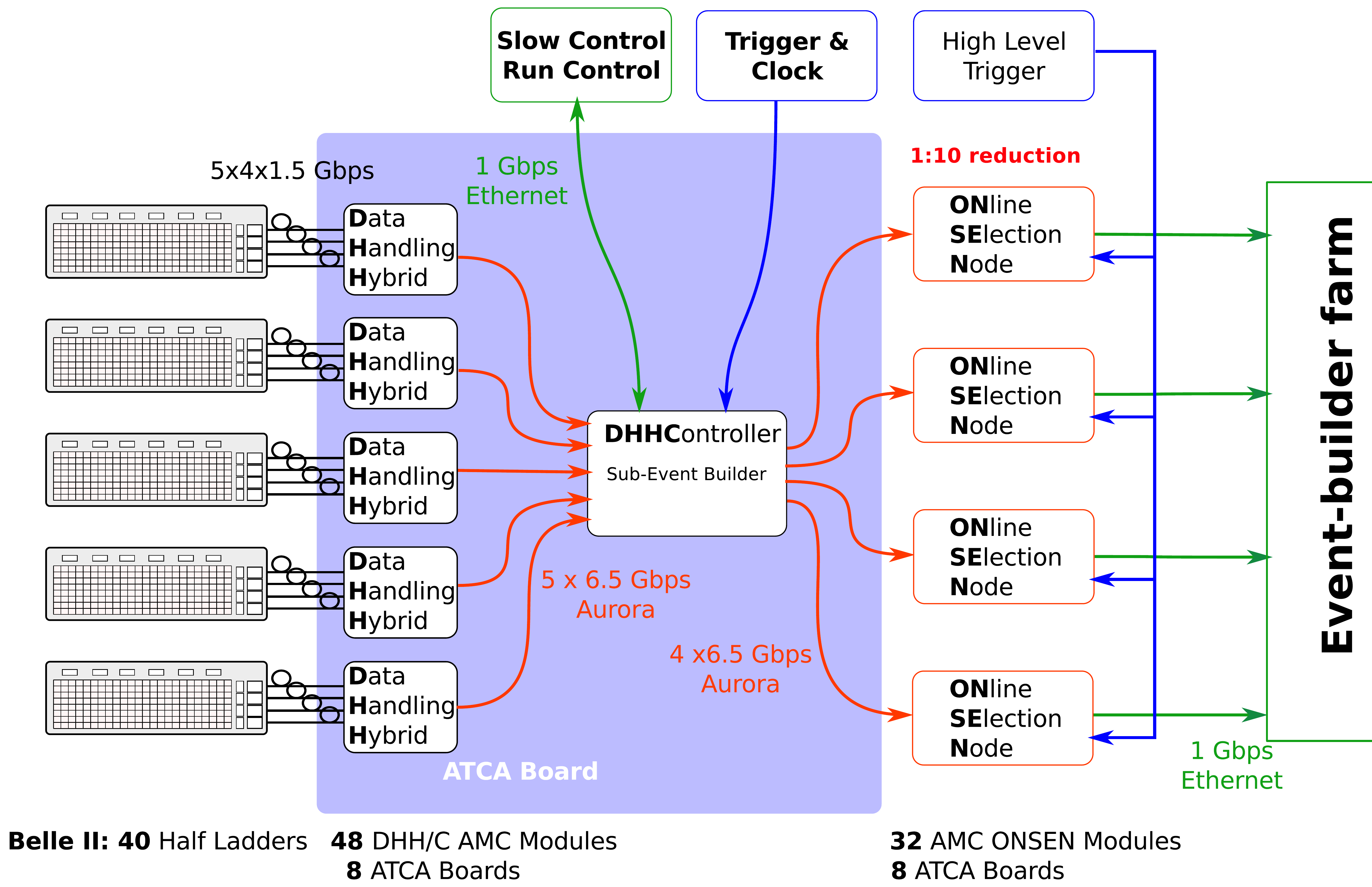}
		\caption{Data read-out chain of the Belle 2 pixel detector}
		\label{fig:pxd-ro-chain}
	\end{figure}

The half ladder forms the read-out unit. 
The DEPFET matrix is operated by three kinds of ASICs bump bonded on the silicon frame of the half ladder: six SwitcherB, four Drain Current Digitizers\,(DCD) and four Data Handling Processors\,(DHP).
The gate and clear gates of the DEPFET matrix are steered by the SwitcherB\,\cite{6154365} ASIC.
These chips are connected in the daisy chain that propagates the externally generated detector read-out sequence.
The gate and clear lines of each four rows in the matrix are controlled by the same SwitcherB channel in a so-called 4-fold read out: four detector rows are digitized in the same read-out cycle.
The read-out cycle consists of activating four detector rows and clearing rows at the end of the cycle.
The period of the read-out cycle is 100\,ns.

The digitization of the drain current is performed in the DCD ASIC\,\cite{dcdb}.
A DCD has 256 8-bit ADC channels with common-mode correction operated at 305\,MHz.
The digitized values are serialized and sent to the DHP ASIC at a total data rate of 20.48\,Gbps.

The DHP ASIC\,\cite{dhp} generates the control sequence for the SwitcherB and synchronizes it with the signal digitization in the DCD.
Initial data processing, e.g. digital common-mode correction, zero suppression, or pedestal subtraction, is performed in the chip.
The DHP can withstand an average detector occupancy up to 3\,\% with negligible data loss.
The processed data is serialized and sent to the data-read-out hardware over a 1.52\,Gbps 8b/10b simplex Aurora channel.
Control of the read-out process is performed by the read-out hardware over four LVDS lines.

All ASICs on the half ladder are connected in the JTAG chain.
The JTAG chain is used for slow control and configuration of the ASICs that is performed by the read-out hardware.

\section{The DHH System}

The read-out hardware of the Belle 2 pixel detector is represented by the Data Handling Hybrid system\,(DHH), which has a two-layer architecture\,(fig.\,\ref{fig:pxd-ro-chain}).
The first-layer modules, the DHHs, are interfaced over two 15\,m Infiniband cables to the front-end electronics on the half ladders.
The modules receive data, perform cluster reconstruction and classification, and provide a JTAG master interface for slow control of the front-end ASICs.
The second layer modules, the DHH Controllers\,(DHHC), are connected to five DHH modules over 6.25\,Gbps Aurora channels. 
A DHHC module collects hit information from DHH modules and performs sub-event building based on the trigger number.
Additionally, a DHHC module provisions interfaces with the trigger and clock distribution system and slow control. 
The synchronization information and ethernet frames are distributed to the DHH modules over high-speed serial links.
Five DHHs and one DHHC module are installed on the ATCA carrier board and form the read-out unit of the system.

\subsection{Hardware}

	\begin{figure}
		\includegraphics[width=3.5in]{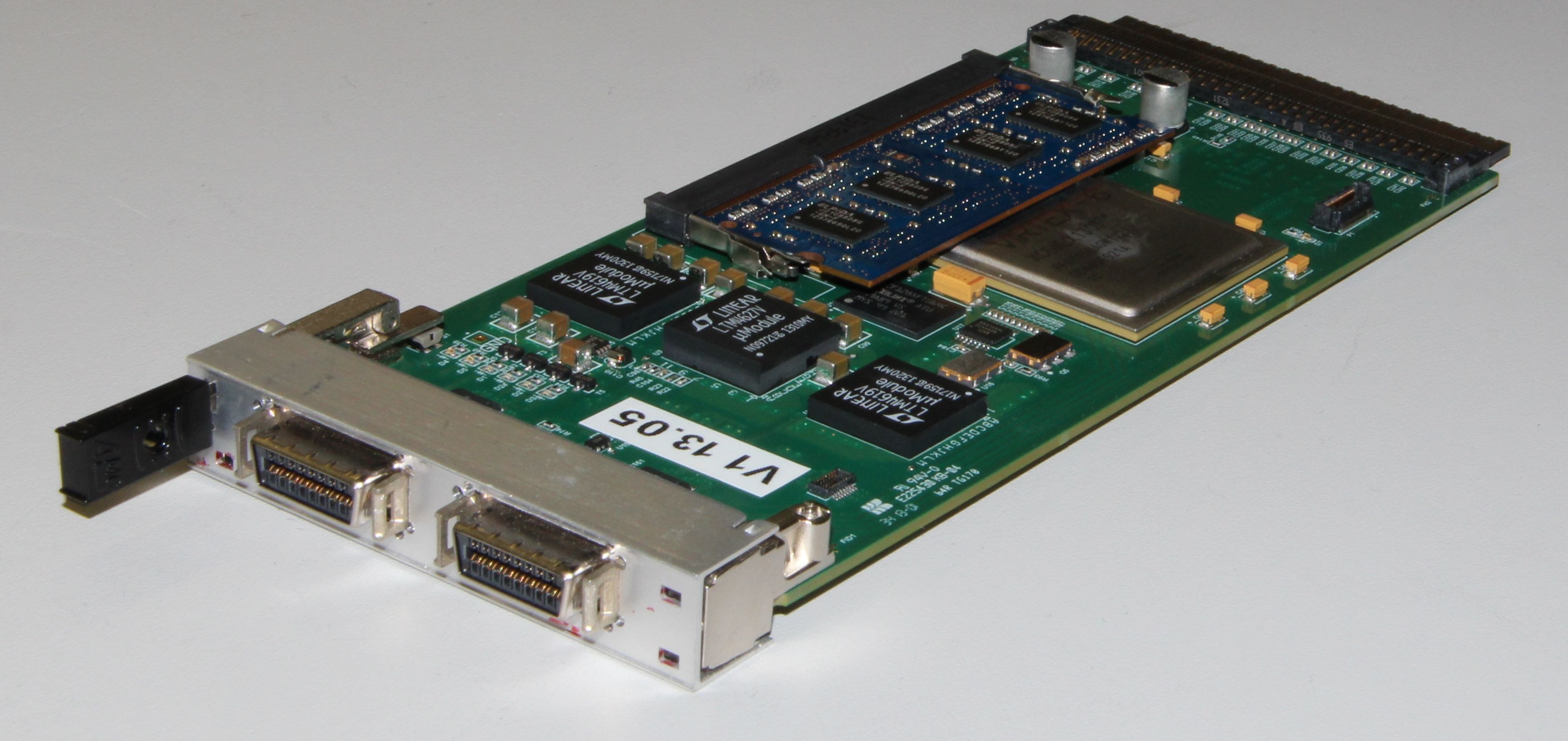}
		\caption{DHH AMC module}
		\label{fig:dhh-amc}
	\end{figure}

The DHH and DHHC modules are built in the AMC form factor and share a hardware design\,(fig.\,\ref{fig:dhh-amc}). 
The main element of the module is a Virtex-6 VLX130T FPGA.
The module is connected to a half ladder using two infiniband connectors.
The DDR3 SODIMM allows up to 4\,GB of memory and a bandwidth up to 6\,GBps.
The configurable clock synthesizer is used as a source of clock signal for stand-alone setup and as a jitter cleaner in setups with an external clock.
There is a designated connector for a current-mirror mezzanine board used for the characterization of the ADCs in the DCD, and a connector for the MMC mezzanine board.

\subsection{System Synchronization}

The pixel detector in the Belle 2 experiment is synchronized with the SuperKEKB accelerator.
The synchronization is done by the Belle 2 Trigger and Time distribution system\,(B2TT)\,\cite{6545397}.
The B2TT system broadcasts a 127.21\,MHz clock that is generated from the accelerator's radio frequency clock of 508.84\,MHz and synchronized to the beam revolution cycle of 100\,kHz\,\cite{1748-0221-7-01-C01028}.
The trigger and synchronization information is distributed source synchronous with the B2TT clock as an 8b/10b encoded serial data stream.

The DHHC synthesizes the 76.33\,MHz clock from the B2TT clock.
The new clock is then distributed to all connected DHH modules and is used as a clock source for the front-end electronics.
The trigger information is propagated synchronously through the custom clock domain crossing module before being distributed to the DHH modules.
This module is programmed with the assumption that one clock is a fraction of the second clock.
Therefore after a defined time period, the phase relation between the clocks will repeat.
Data are written to a register at the beginning of this period in one clock domain and read from it at the end of the period in the other.
This reduces the time resolution but is well within the trigger interval of 190\,ns \cite{6545397}.

The 76.33\,MHz clock and data are distributed synchronously from the DHHC to DHH modules over a dedicated high-speed data link.
The phase information of the source clock is preserved by using 8b/10b encoding, which makes clock recovery on the receiver side possible.
The bypassing of the elastic buffers in transmitter and receiver also fixes the latency for the transmission of the trigger information.
The recovered clock on the receiver side has a fixed phase relation to the source clock.

\subsection{Data Processing}

	\begin{figure}
		\includegraphics[width=3.7in]{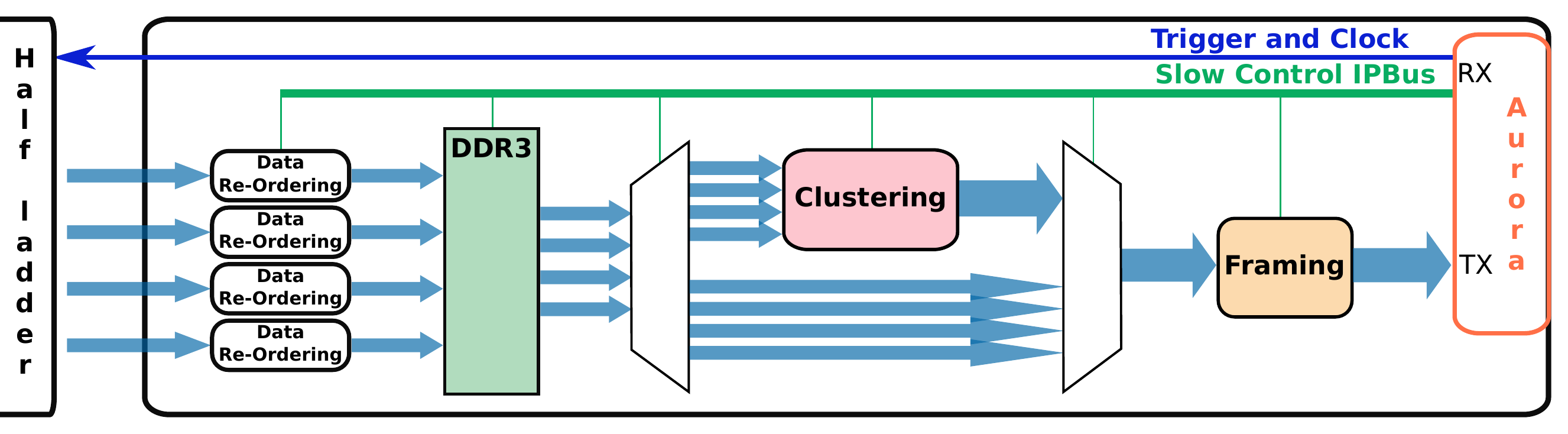}
		\caption{Data flow in the DHH}
		\label{fig:dhh_data_flow}
	\end{figure}

The data flow in the DHH is shown in fig.\,\ref{fig:dhh_data_flow}: data are received from DHPs by four independent Aurora cores.
The DHP data is divided only in DEPFET frames when the data pointer in the DHP memory transitions from the last row to the first row.
The data that belong to overlapping triggers are disentangled in the DHH.

The algorithm consists of the job allocator, data-storage modules, and data reader.
The number of the data-storage modules defines the maximum number of overlapping triggers that the DHH can process.

The job allocator uses a round-robin algorithm to activate data storage modules and provides information about the trigger.
This information includes the trigger number, the number of the first row that will be processed by the DHP for this trigger, and the expected DHP frame ID.
Because DHH and DHP share the same clock, the number of the first row is calculated in the DHH and corresponds to the data pointer in the DHP.
The activated data-storage module analyzes the data stream to filter exactly one DEPFET frame from the DHP data stream and stores data frames in the FIFO.
Finally, the data reader also uses a round-robin algorithm to read finished events from data-storage modules.

Another problem is addressed in the data-storage module.
If a DEPFET event is received in two DHP frames, than the row number inside of the event jumps between two frames.
The correction of this condition is implemented in the module.
The frames are stored in separate FIFOs.
Then the frames are read in reverse order and merged to a single frame.
Finally, the single DHP frame is stored with the trigger tag in the DDR3 memory.

\begin{figure}
		\includegraphics[width=3.8in]{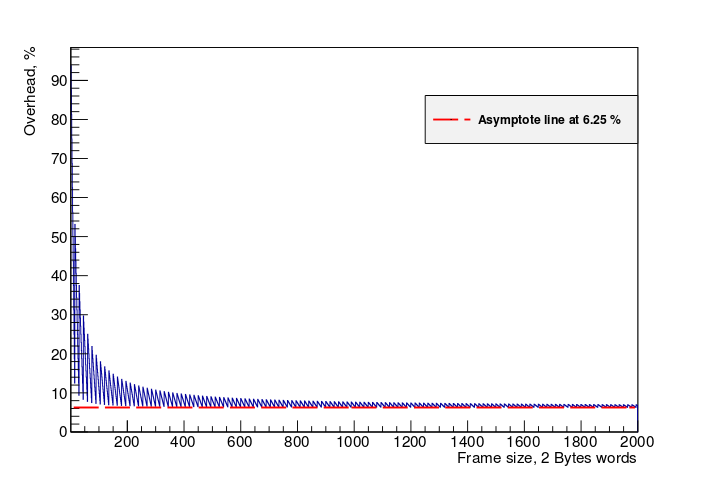}
		\caption{Overhead of the external FIFO core (simulation)}
		\label{fig:ddr3_fifo_overhead}
	\end{figure}

The DDR3 memory provides four FIFO-like LocalLink interfaces for storing detector frames.
The custom core is built around the Xilinx MIG core to share the same DDR3 memory unit.
The data are written into an intermediate FIFO that keeps data before access to the memory interface is granted by the memory arbiter module.
The arbiter divides single address space of the DDR3 memory into several memory regions that are treated as ring buffers.
The arbiter maintains a set of pointers for each ring buffer.
The size of the ring buffers can be dynamically changed in the slow control registers.
If data are present in the intermediate FIFO, then the next time the arbiter activates this memory region, a block of data is written to the DDR3 memory.
The read process is similar to the write: the arbiter reads data from the ring buffer and writes it into the intermediate FIFO, where it is decoded and prepared for the LocalLink interface.
The overhead of the algorithm goes asymptotically to 6.25\,\% since 16 bits of the 256-bit vector in the MIG interface are used for service information.
The last vector of the frame may not be filled completely and therefore the overhead also oscillates with the frame size (fig.\,\ref{fig:ddr3_fifo_overhead}).

The buffered data are then optionally processed by the cluster-recovery algorithm.
The clustering of the data is beneficial for the effective implementation of the data-reduction algorithms.
The clustering algorithm takes advantage of the fact that the hits are already ordered in incremental row number sequence.
The core of the algorithm is implemented as a chain of finite state machines, representing one detector row. 
Each state machine processes hits of two neighboring columns. 
Therefore the algorithm has 32 state machines for 64 detector columns corresponding to the number of columns handled by one DHP chip.
The clustering process receives one hit per clock.
The state machine checks the status of its direct neighbours and if one of the neighbours is active, then the state machine takes its cluster number.
Otherwise the next free cluster number is used.
When there are two active neighbours, the clusters are merged and the state machine assigns the smallest cluster number to the hit.
To remap preliminary cluster numbers to final ones there is a table that stores the new cluster number at the address of the old cluster number.
After the preliminary cluster numbers have been assigned to all hits, a remapping of the cluster numbers is performed in the order of pixels stored in the hit FIFO. 

Since only the smallest cluster numbers are propagated, the cluster number remapping requires at most two look-ups to the table to find the true cluster number.
Once the true cluster number is found, the value is written into the look-up table to speed up the following remapping.
Since the remapping algorithm is deterministic and the hits are received in the correct order, the data stream can be processed in a pipeline.
Finally, four data streams are merged together by remapping clusters that are located on the frame border.

The verification of the algorithm was performed in firmware and in software in parallel.
The data in the pixel detector data format were generated by the software running on a dedicated PC.
Then the data were loaded into FPGA over Ethernet and sent to the software for verification.
The clustered data were downloaded from FPGA and compared with the results of the cluster recovery in hardware.
This test proved the ability of the cluster recovery algorithm in FPGA to recover the clusters in the data stream correctly.

The data ready to be sent downstream are formated into DHH frames that provide service information on the data type and event number.
The DHH frames are then sent to the DHHC module for sub-event building.


	\begin{figure}
		\includegraphics[width=3.5in]{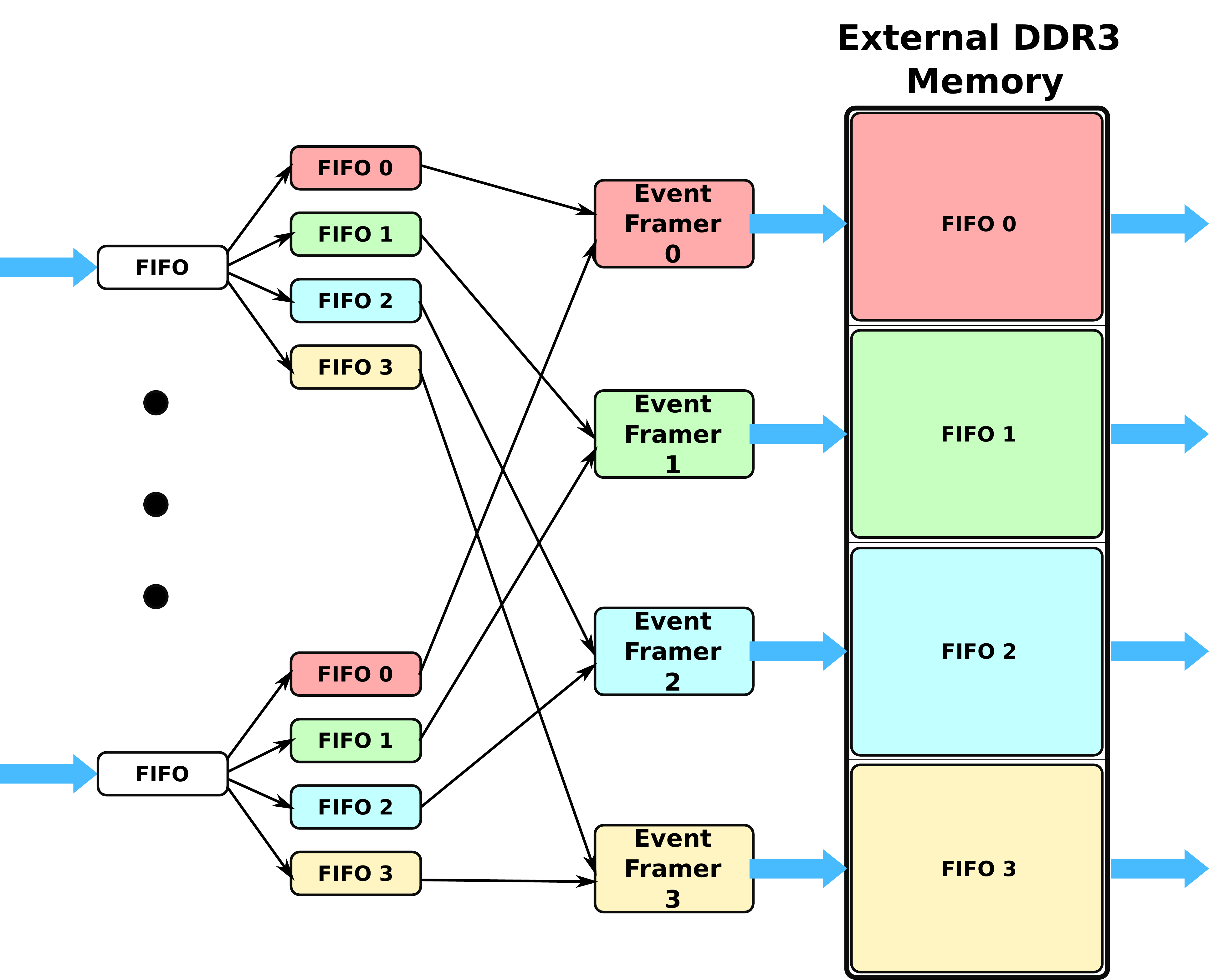}
		\caption{Sub-event builder algorithm}
		\label{fig:sub-evb}
	\end{figure}

The DHHC is used to perform online sub-event building in firmware.
The data flow in the sub-event building algorithm is shown in fig.\,\ref{fig:sub-evb}: the received frames are buffered in an internal FIFO and then written into one of the four intermediate FIFOs that correspond to the four outgoing links.
The active intermediate FIFO is determined by a round-robin algorithm.
Overflow of the buffering FIFO is prevented by using the Native Flow Control feature of the Aurora protocol.
This feature allows us to suspend data transmission over the Aurora link until the fill level of the FIFO falls under a pre-defined threshold.
During this time period the data are buffered in the external memory of the DHH modules.

The frames stored in the intermediate FIFOs are then read out by framing state machines.
The order of the FIFOs that are read out is pre-determined and starts with a FIFO that is connected to an active incoming link and has the smallest index.
These state machines enclose frames with the same event number by a header and a trailer frame, thereby building a sub-event that contains information from up to five detector modules.
Finally, the complete sub-events are written into a large FIFO that is implemented in the external memory.
The data from the external FIFOs are directly sent to outgoing links.

Since the data frames already contain event number information, the consistency of the sub-event is directly checked in the framing state machines.
The sub-event-builder module also provides the possibility to mask incoming and outgoing channels.
In this case, the round-robin algorithm is altered to bypass inactive channels.

\subsection{Slow Control}

	\begin{figure}
		\includegraphics[width=3.5in]{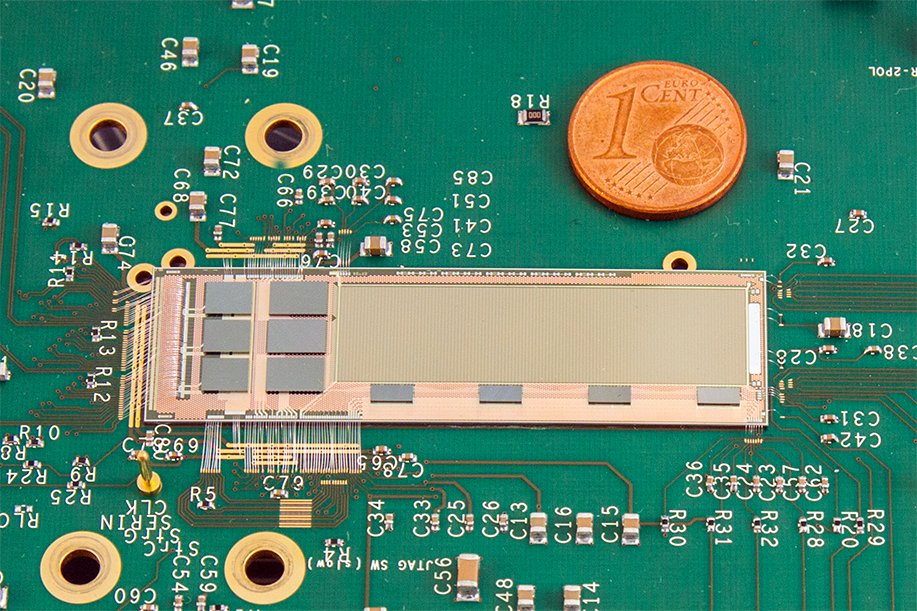}
		\caption{DEPFET module bonded on the Hybrid6 board. Courtesy of \mbox{M. Schnell}}
		\label{fig:hybrid6_matrix}
	\end{figure}

The slow control of the system is implemented as an abstraction layer between high-level system-control algorithms in EPICS and low-level hardware registers.
The DHH is controlled directly over Ethernet using a UDP-based protocol IPBus.
The front-end ASICs are connected into a JTAG chain and are controlled by a hardware master in the DHH.
A middleware library was developed that translates IPBus and JTAG protocols for the high-level EPICS slow-control network.

The ATCA carrier board provides only a single Ethernet connection for the whole read-out system.
All modules on the carrier board are accessible by a unique IP address.
This is done by implementing a simple Ethernet hub in the FPGA logic of the DHHC to share this connection with the DHH modules.
The hub broadcasts all received Ethernet frames to the DHH modules and to the IPBus client on the DHHC.
The Ethernet frames are transmitted over the same Aurora link that is used to read out data from the DHH.
The replies are multiplexed with the data stream in the DHH.
The multiplexer assigns high priority to the data and low priority to the Ethernet stream.
The received replies on the DHHC are then transmitted into the slow-control Ethernet network.

The JTAG master is implemented as a two-level system.
The hardware master in the DHH consists of a state machine that executes external commands.
The control registers of the hardware master are mapped to the IPBus registers.
The software master is built as an EPICS driver using the \mbox{asynPortDriver} class.
The main task of this master is generation of the JTAG commands and transmission of the commands to the hardware master on the DHH over IPBus.
The software maintains not only the knowledge of the JTAG registers available in the ASICs, but also of the bit fields inside of the registers.
Every bit field that controls a specific function in an ASIC is connected to an EPICS register in the software.
If a JTAG register access is scheduled, the software constructs the bitstream using the cached values of the bit fields and writes the bistream commands into the corresponding DHH module over IPBus.

	\begin{figure}
		\includegraphics[width=3.5in]{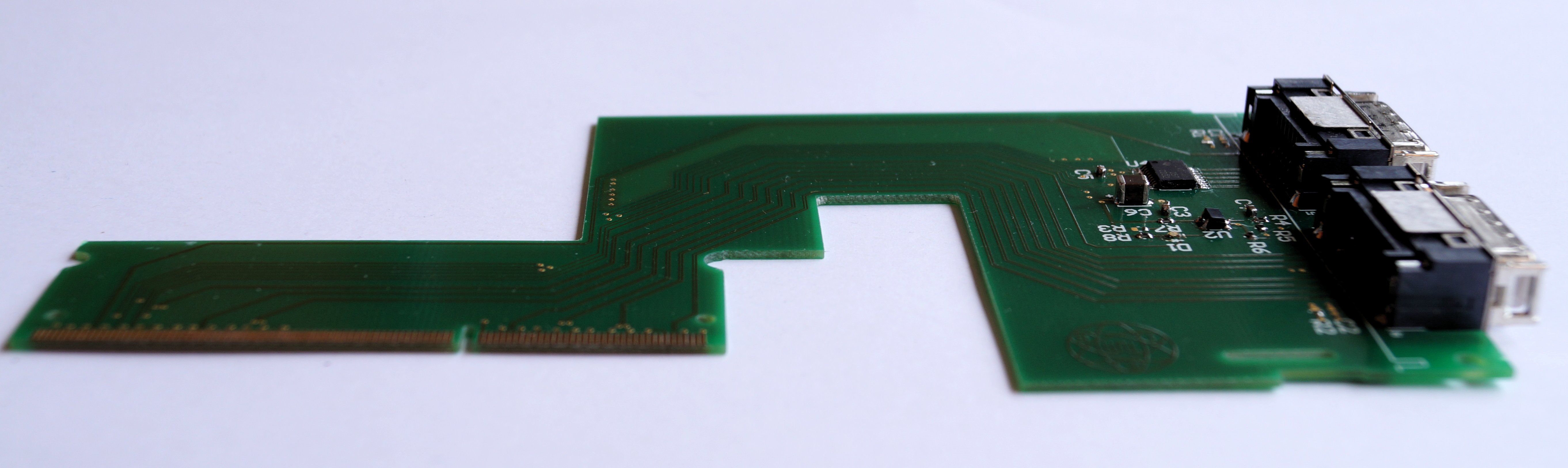}
		\caption{SODIMM Adapter Board}
		\label{fig:sodimm-ab}
	\end{figure}


\section{System Test at DESY in January 2014}

The first test of the full system was performed at DESY in January 2014.
A beam of electrons and positrons produced from the bremsstrahlung of the DESY II accelerator was used for the test.
The goal of the campaign was to test the integration of the data-acquisition and online-data-reduction systems of the Belle 2 vertex detector.
The test was performed on a detector prototype consisting of one layer of the pixel detector and four layers of the double-sided silicon strip detector\,\cite{Friedl201383}.
While the detector provides enough planes for track reconstruction, the AIDA telescope was used to estimate the performance of the detector.
Three planes of the telescope were installed before the detector and another three planes were installed downstream.
The detector and the telescope were installed inside of the superconducting magnet PCMAG that allowed us to study detector performance in a range of magnetic fields up to 1\,T.

\subsection{Pixel Detector Module}

The pixel detector module in the beam test, the Hybrid 6 board\,(fig.\,\ref{fig:hybrid6_matrix}), has a DEPFET matrix of 480x192 pixels with the pixel area of 50x75\,$\mu$m$^2$.
Three pairs of DHP and DCD ASICs and four SwitcherB ASICs are installed on the module.
The main difference of this module compared with the Belle 2 design is the control of the SwitcherB ASICs: the ASICs are not controlled by the DHP, but by the DHH instead.
The Hybrid 6 board offers two additional Infiniband connectors for the SwitcherB control signals.

The DHH is extended by the adapter board for the SODIMM slot\,(fig.\,\ref{fig:sodimm-ab}).
This board carries two Infiniband ports and a single-ended-to-LVDS signal converter.
The control functionality of the DHH requires the extension of the firmware with two additional modules: the sequencer and the JTAG player.

The sequencer creates the sequence of the signals for SwitcherB ASICs that controls rolling shutter mode operation of the detector.
The sequencer is implemented as a dual port memory.
The sequence is programmed by the slow control system and read continuously until the end of the sequence in the memory is reached or a frame synchronization signal is received.
Then the read pointer resets and the frame cycle repeats.

The implementation of the switcher slow control over JTAG imitates the main JTAG master.
The second JTAG player in firmware is controlled by the second instance of the software master, which maps switcher registers to EPICS registers.
This eliminates the need for modification of the GUI and high-level logic.

\subsection{Detector Read-Out Chain}

	\begin{figure}
		\includegraphics[width=3.5in]{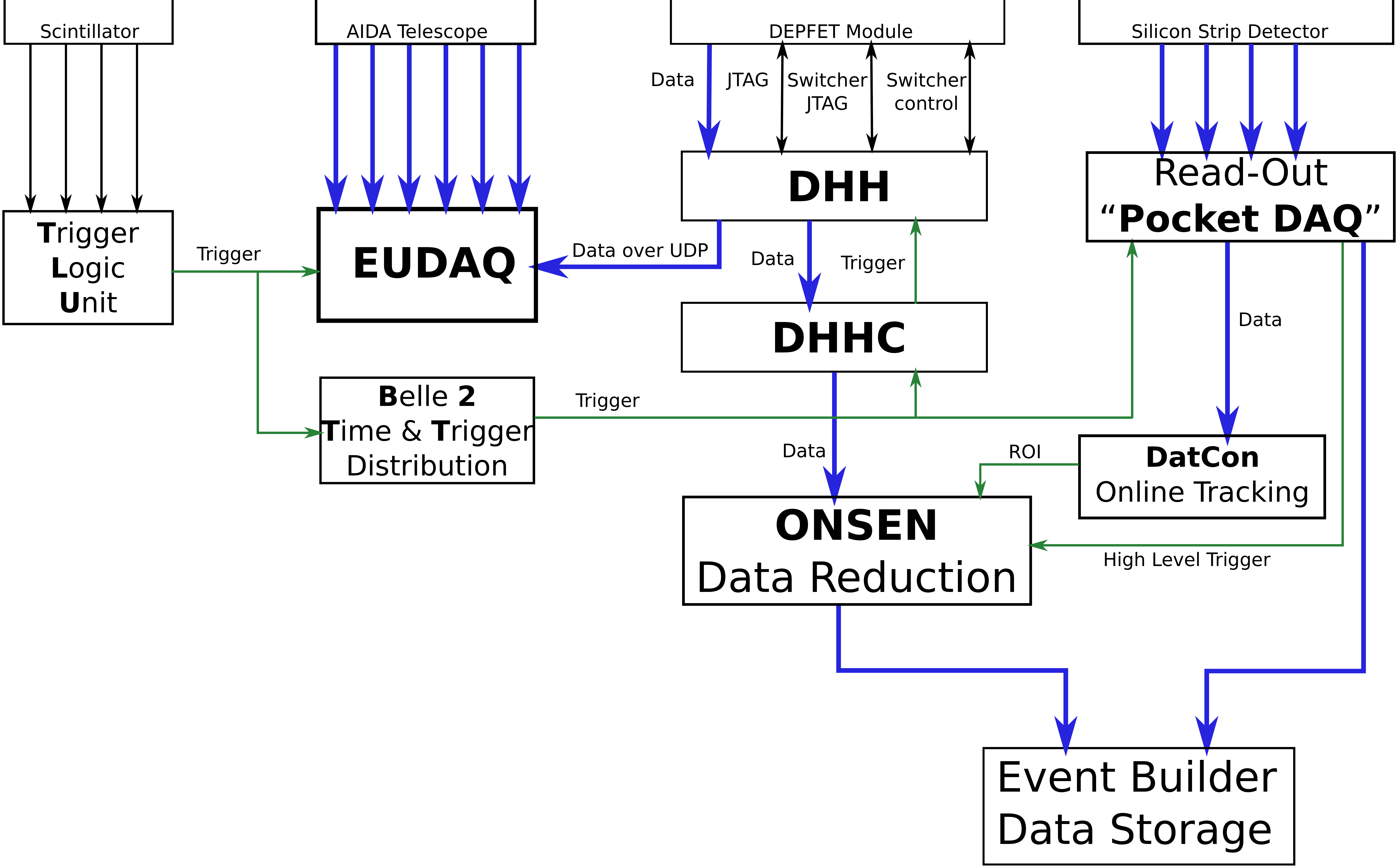}
		\caption{Read-out chain at the beam test setup}
		\label{fig:readout_chain_jan2014}
	\end{figure}

The data read-out chain of the detectors is shown in fig.\,\ref{fig:readout_chain_jan2014}.
The fast trigger signal generated by scintillators is transmitted to the Trigger Logic Unit\,(TLU) to be used as a level 1 trigger.
The TLU assigns consecutive event numbers to the triggers and transmits them to the telescope and to the B2TT system.
The B2TT system passes along this information and triggers the read-out of the pixel detector and silicon vertex detector.

Once the trigger is received by the DHHC module, it is transmitted to the DHH module, where the trigger signal for the front-end electronics is generated.
In response to the trigger signal, the DHP chips send up to two data frames in the zero-suppressed format.
Then the frames are formated into DHH format and sent to the DHHC.
The DHHC formats data in the DHHC format and sends them to the Online Selection Node\,(ONSEN)\,\cite{6615996}.
A copy of the DHH data is also sent over Ethernet to the standalone DAQ PC.
On the standalone DAQ PC the data are monitored by the data quality monitor and forwarded to the EUDAQ PC to be merged with the telescope data.

The ONSEN system is designed to perform online data reduction by using the information from the outer detectors of Belle 2.
The data from outer detectors are used to generate Regions of Interest\,(ROI), the intersection points of the particle tracks with the planes of the pixel detector, by extrapolating the tracks down to the interaction point.
The ROIs are received by ONSEN from two sources: the FPGA-based silicon-strip-detector-only track finder Data Concentrator\,(DATCON) and software-based high-level trigger.
Both ROI sources were used in the beam test although they used data from the same detector.
The pixel data were filtered in the ONSEN by checking the data against ROIs.
The filtered data were then sent to the event builder farm, which merged data streams from the pixel detector and silicon strip detector.

\subsection{System Performance at the Beam Test}

\begin{figure}
		\includegraphics[width=3.5in]{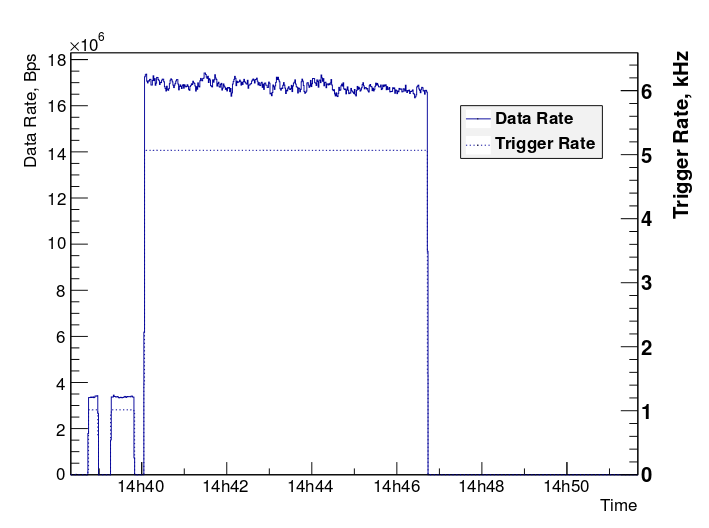}
		\caption{Data and trigger rates test at the beam test setup}
		\label{fig:data_trigger_rate}
	\end{figure}

A dry run was performed on the DHH--DHHC setup to test the performance of the system before declaring the system operational.
An artificial trigger with a defined frequency was generated by the B2TT system, and the data were read out from the DEPFET module and sent to the ONSEN module.
No high-level trigger was generated during the dry run and therefore the run duration was limited by the capacity of the memory in the ONSEN system.
The system ran at a trigger rate of 5\,kHz with a continuous data stream of 17\,MBps for 6\,minutes without errors (fig.\,\ref{fig:data_trigger_rate}).

The read-out system was in continuous operation for seven days with the average trigger rate of 300\,Hz.
The rate of the data stream generated by the detector averaged around 200\,KBps.
During the beam test, approximately 60 million good events were recorded by the system.

\section{Conclusions}

The read-out system for the Belle 2 pixel detector was developed and integrated into the EPICS slow control and the pixel detector read-out chain.
Its key features were successfully tested in a beam test at DESY as the integral part of the silicon vertex detector read-out chain.

\section{Acknowledgements}

The research leading to these results has received funding from the European Commission under the FP7 Research Infrastructures project AIDA, grant agreement no. 262025.
This work is supported by the German Ministry of Education and Research, Excellence Cluster Universe and the Maier-Leibnitz-Laboratory.

\bibliographystyle{IEEEtran}
\bibliography{nss}

\end{document}